\documentclass[twocolumn,aip,rsi,reprint]{revtex4-1}
\usepackage[latin9]{inputenc}
\usepackage{xcolor}
\usepackage{pdfcolmk}
\usepackage{amsmath}
\usepackage{amssymb}
\usepackage{graphicx}
\PassOptionsToPackage{normalem}{ulem}
\usepackage{ulem}
\usepackage[unicode=true,pdfusetitle,
 bookmarks=false,
 breaklinks=false,pdfborder={0 0 0},pdfborderstyle={},backref=false,colorlinks=true]
 {hyperref}
\hypersetup{
 colorlinks,linkcolor=red,citecolor=blue}

\makeatletter

\providecolor{lyxadded}{rgb}{0,0,1}
\providecolor{lyxdeleted}{rgb}{1,0,0}

\usepackage{amsfonts}

\makeatother

\begin{document}

\title{Phase sensitive imaging of 10 GHz vibrations in an AlN microdisk
resonator}

\author{Zhen Shen, Xu Han, Chang-Ling Zou, and Hong X. Tang}
\email{hong.tang@yale.edu}

\address{Department of Electrical Engineering, Yale University, New Haven,
Connecticut 06511, USA}
\begin{abstract}
We demonstrate a high frequency phase-sensitive heterodyne vibrometer,
operating up to 10 GHz. Using this heterodyne vibrometer, the amplitude
and phase fields of the fundamental thickness mode, the radial fundamental
and the 2nd-order modes of an AlN optomechanical microdisk resonator
are mapped with a displacement sensitivity of {\normalsize{}around
$0.36\:\textrm{pm}/\sqrt{\textrm{Hz}}$}. The simultaneous amplitude
and phase measurement allows precise mode identification and characterization.
The recorded modal frequencies and profiles are consistent with numerical
simulations. This vibrometer will be of great significance for the
development of high frequency mechanical devices.
\end{abstract}
\maketitle
Electromechanical and optomechanical resonators have played important
roles in a variety of quantum and classical applications \citep{Appli1,Appli2,Appli3},
including ground-state cooling of mechanical resonators \citep{Cool1,Cool2,Cool3},
strong coupling between mechanical oscillator and microwave/optical
resonators \citep{strong1,strong2,strong3}, efficient conversion
between photons at vastly different frequencies \citep{Conversion1,Conversion2},
and non-reciprocal optomechanical devices \citep{Nonreciprocity1,Nonreciprocity2,Nonreciprocity3}.
Particularly, high frequency mechanical devices above 10 GHz are desirable
for advancing mechanical quantum devices because they require less
stringent refrigeration conditions for reaching quantum mechanical
ground state and their frequency are matched to superconducting quantum
circuits \citep{10GHzImprotant}. A 10 GHz mechanical oscillator can
be cooled to ground state by direct dilution refrigeration without
the necessity of active cooling techniques. Furthermore, the extension
of acoustic resonators to microwave X-band and beyond is important
for next generation wireless communications \citep{book1,communication1,MoLi1,MoLi2}.

The characterization of vibrational modes of electro/optomechanical
devices, including identification of the vibration frequencies, profiles,
mode losses and so on, is important for validating their design, operation
and performance. Scanning optical vibrometer enable high-resolution,
non-invasive mapping of the vibration pattern of acoustic devices\citep{Vib1SAW,Vib2thin,Vib3thin,Timeresolved1,VibConfocal,Vib4Canti,Vib5drump,Vib6thin,Timeresolved2,Vib7Canti,Vib8thin,Vib9drump,VibSiC,Vibwaveguide,Vibring}.
It also offers complete quadrature information rather than pure intensity
response. Optical vibrometers have proven their strength in the researches
and applications of microacoustic components, including thin film
resonators \citep{Vib2thin,Vib3thin,Vib6thin,Timeresolved2,Vib8thin},
piezoelectric microcantilevers \citep{Vib4Canti,Vib7Canti}, silicon
carbide microdisk resonators \citep{VibSiC}, surface acoustic resonators
and waveguides on a chip \citep{Vibwaveguide,Vibring}. Optical pump-probe
technique, which requires high-performance pulsed laser, allows the
time-domain response of mechanical oscillators to be measured \citep{Timeresolved1,Timeresolved2,Timeresolved3,Timeresolved4}.
Frequency-domain measurements with continuous laser, as a complementary
method, is often used to characterize mechanical resonators with high
quality ($Q$) factors. However, the frequencies of the vibration
modes mapped by this method are limited to a few gigahertz \citep{Vib2thin,Vib3thin,Vib6thin,Vib8thin}.
Here we demonstrate a scanning heterodyne vibrometer for realizing
simultaneous absolute amplitude and phase imaging of vibration up
to 10 GHz-range in an AlN microdisk resonator with a displacement
sensitivity of $0.36\:\textrm{pm}/\sqrt{\textrm{Hz}}$ for out-of-plane
motion detection. Measurements on the fundamental thickness mode,
the fundamental radial mode and the 2nd-order radial modes of an AlN
microdisk provide phase-sensitive modal visualization that is consistent
with our numerical simulation. 

\begin{figure}
\includegraphics[bb=198bp 103bp 519bp 438bp,clip,width=1\columnwidth]{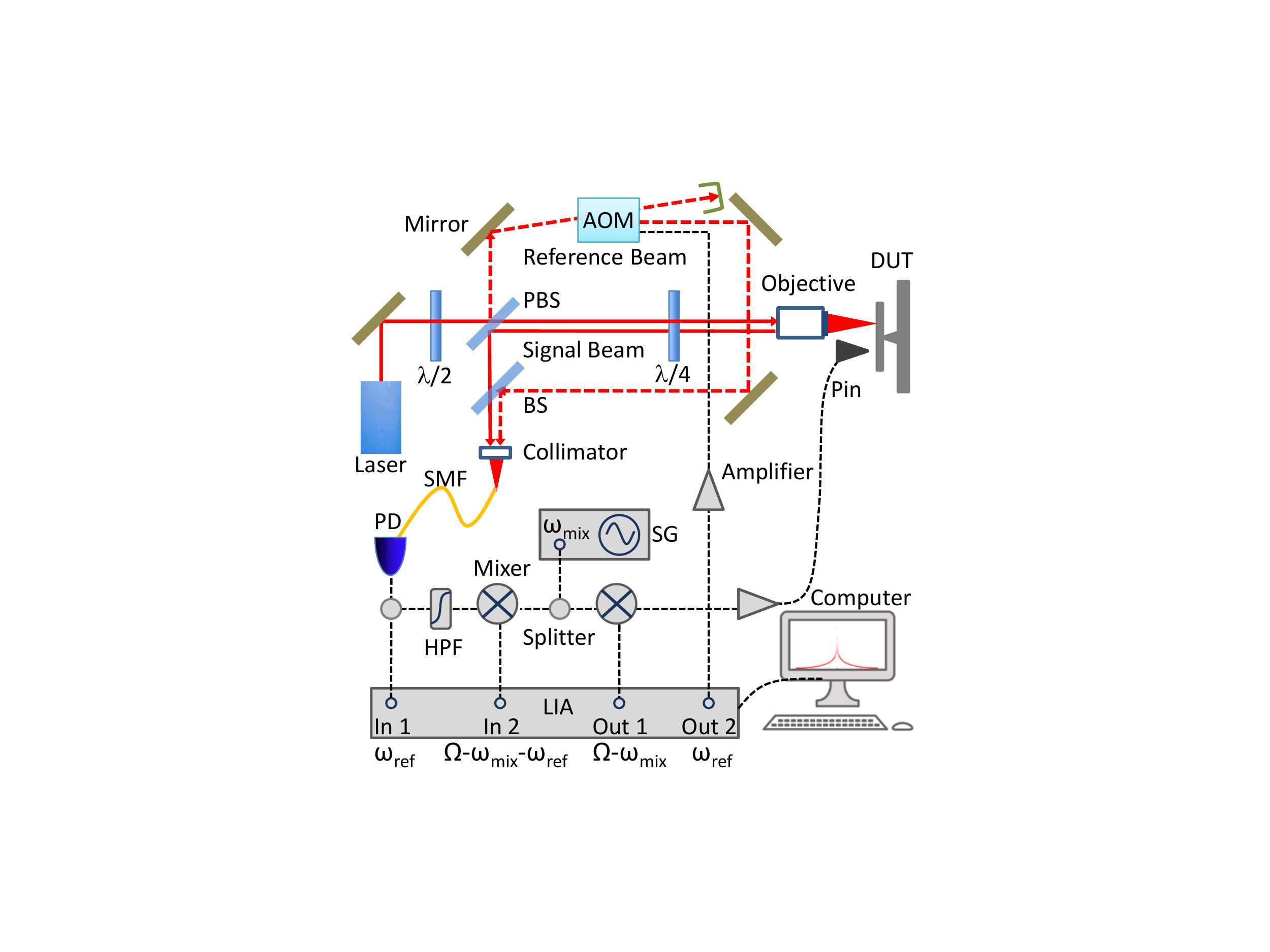}\caption{Schematic layout of the heterodyne vibrometer. AOM: acousto-optic
modulator, PBS: polarizing beam splitter, BS: beam splitter, DUT:
device under test, Pin: dipolar pin antenna actuator, SMF: single
mode fiber, PD: photodetector, HPF: high-pass filter, SG: signal generator,
LIA: lock-in amplifier.}

\label{fig1}
\end{figure}

The vibrometer derives the vibration from the modulated light signal
by a heterodyne interoferometer. The experimental setup is shown in
Fig.$\,$\ref{fig1}. A similar setup has also been realized and described
in detail by Martinussen \emph{et al.} \citep{Vib5drump}, Kokkonen
and Kaivola \citep{Vib6thin}, and Leirset \emph{et al.} \citep{Vib9drump}.
A linearly polarized 633 nm HeNe laser is split into two paths - the
reference beam and the probe beam - by a polarizing beam splitter
(PBS). A half-wave plate ($\lambda/2$) is used to tune the power
ratio between the reference and the probe beams. The frequency of
the reference beam is shifted from the original input laser frequency
$\omega$ to $\omega+\omega_{\textrm{ref}}$ by an acousto-optic modulator
(AOM). The probe beam is focused perpendicularly onto the sample with
a microscope objective (NA $=0.55$) and reflected back to the PBS.
The polarization of the reflected probe signal is rotated by 90 degrees
because of the double pass through the $\lambda/4$ plate, which leads
to the same polarization as the frequency-shifted reference light.
Then the motion-modulated probe and reference light are combined with
the beam splitter and sent into a photodetector (PD) through a single
mode fiber. The fiber acts as a confocal pinhole to increase the lateral
resolution \citep{VibConfocal}. The theoretical resolution limit
in our setup is around 600 nm ($0.5\lambda/\textrm{NA}$) as described
in Ref.$\,$24. 

To analyze the heterodyne signal, we consider the sample surface vibrating
sinusoidally as $Z_{0}\mathrm{sin}(\Omega t+\phi)$ in the z-direction
(perpendicular to the surface of sample), where $Z_{0}$, $\Omega$
and $\phi$ are the amplitude, angular frequency and phase of the
vibration, respectively. The probe light field detected by the PD
is $E_{p}=A_{1}\mathrm{cos}[\omega t+\phi_{1}+2kZ_{0}sin(\Omega t+\phi)]$
and the reference light field is $E_{\textrm{ref}}=A_{2}\mathrm{cos}[(\omega+\omega_{\textrm{ref}})t+\phi_{2}]$,
where $A_{1}$ ($A_{2}$) and $\phi_{1}$ ($\phi_{2}$) denote the
amplitudes and phases of the probe (reference) field, and $k=2\pi/\lambda$
is the optical wavenumber. When $Z_{0}\ll\lambda$, the $E_{\textrm{p}}$
can be expanded to the first order of $Z_{0}$ as
\begin{eqnarray}
E_{\textrm{p}} & \approx & A_{1}\mathrm{cos}(\omega t+\phi_{1})+A_{1}kZ_{0}\mathrm{cos}[(\omega+\Omega)t+\phi_{1}+\phi)]\nonumber \\
 &  & -A_{1}kZ_{0}\mathrm{cos}[(\omega-\Omega)t+\phi_{1}-\phi)]
\end{eqnarray}
 The PD response to the intensity of light field is
\begin{eqnarray}
I(t) & = & \alpha\left|E_{\mathrm{p}}+E_{\mathrm{ref}}\right|^{2}\nonumber \\
 & \approx & \alpha\{A_{1}A_{2}\mathrm{cos}(\omega_{\mathrm{ref}}t+\phi_{2}-\phi_{1})\nonumber \\
 &  & +A_{1}A_{2}kZ_{0}\mathrm{cos}[(\Omega-\omega_{\textrm{ref}})t+\phi+\phi_{1}-\phi_{2})]\}
\end{eqnarray}
where $\alpha$ is the responsivity of the PD. Here, we have dropped
the second-order term of $Z_{0}^{2}$ and also filtered out the component
oscillating at high frequency $\Omega+\omega_{\textrm{ref}}$. When
the lock-in amplifier (LIA) demodulates at $\omega_{\textrm{ref}}/2\pi$
and $\omega_{\textrm{sig}}/2\pi=(\Omega-\omega_{\textrm{ref}})/2\pi$
simultaneously, the amplitude $R_{\textrm{ref}}=A_{1}A_{2}$ ($R_{\textrm{sig}}=kZ_{0}R_{\mathrm{\textrm{ref}}}$)
and phase $\theta_{\textrm{ref}}=\phi_{2}-\phi_{1}$ ($\theta_{\textrm{sig}}=\phi+\phi_{1}-\phi_{2}$)
of the reference (signal) component are determined. Then, the amplitude
and the phase of the surface vibration can be calculated as $Z_{0}=\frac{1}{k}\frac{R_{\textrm{sig}}}{R_{\textrm{ref}}}$,
$\phi=\theta_{\textrm{sig}}+\theta_{\textrm{ref}}$.

\begin{figure}
\includegraphics[clip,width=1\columnwidth]{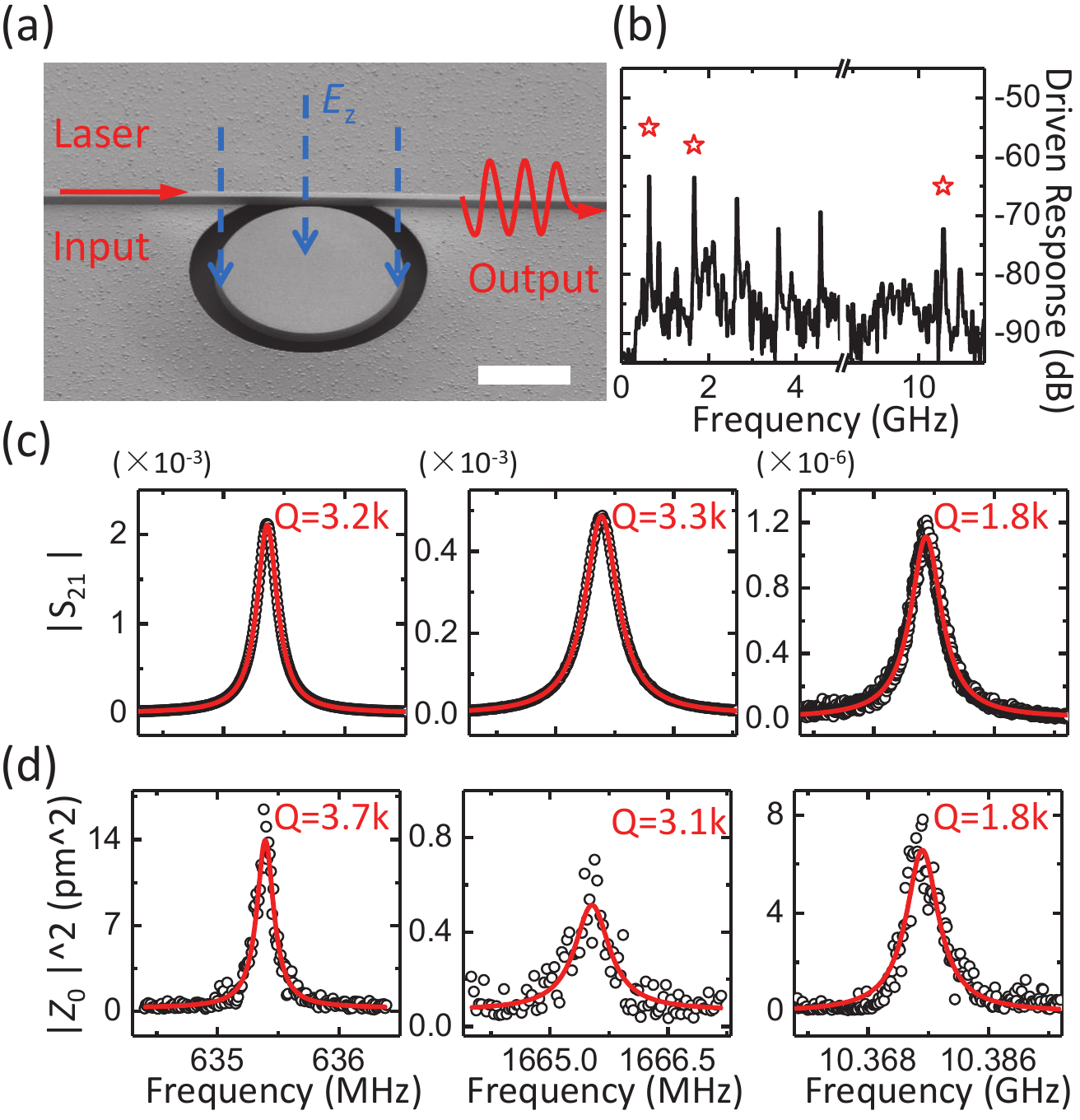}

\caption{(a) SEM image of a microdisk resonator of $5.5\:\textrm{\ensuremath{\mu}m}$
in radius coupled with optical waveguide. The scale bar is $5\:\mu\textrm{m}$.(b)
Piezo-electric driven response of the disk transduced by the integrated
waveguide. The red stars mark the resonance peaks of the fundamental
and the second-order radial modes and the fundamental thickness mode.
(c) Zoom-in spectra of the three modes at 635.4 MHz, 1665.5 MHz and
10.37 GHz. The $Q$ factors are fitted to be 3200, 3300 and 1800,
respectively. (d) The corresponding resonances measured using the
heterodyne vibrometer. }

\label{fig2}
\end{figure}

In our experiment, the frequency of signal $\omega_{\textrm{sig}}/2\pi$
should be mixed with a signal generator (SG) to a lower frequency
that falls within the frequency operation range of the lock-in amplifier
LIA (Zurich instruments UHFLI), i.e., less than 600 MHz. Therefore,
the detector signal from the PD (Newport, 1580-A, 12 GHz) is split
by a power splitter with one of the outputs sent through a high-pass
filter (HPF) to filter out the reference component, then mixed down
to the new frequency $\omega_{\textrm{sig}}^{\prime}=\Omega-\omega_{\textrm{ref}}-\omega_{\textrm{mix}}$,
where $\omega_{\textrm{mix}}$ denotes the operating frequency of
the SG. The mixer output is fed into the LIA and demodulated at $\omega_{\textrm{sig}}^{\prime}/2\pi$.
The amplitude of demodulation signal is $R_{\textrm{sig}}^{\prime}=\xi R_{\textrm{sig}}$,
where $\xi$ is a conversion factor determined by the conversion loss
of mixer, insertion loss of HPF and coaxial cables, and imbalance
of the power splitter. It should be pointed out that this conversion
factor is non-constant when we measure different mechanical modes
in the AlN microdisk due to the frequency-dependent response of electronic
components. For the measurement at around 10.4 GHz, 1.67 GHz and 635
MHz as will be described below, $\xi$ is around $1/4.87$, $1/2.85$
and $1$, respectively. The signal output 1 of LIA is set at $\Omega-\omega_{\textrm{mix}}$
and mixed with the SG to actuate the device using a dipolar pin antenna.
The signal output 2 of LIA is used to drive the AOM. The SG is clock-referenced
to the LIA to ensure that they are phase locked. The device under
test is placed on a linear piezo stage (Micronix PPS-20) with lateral
resolution of $10\:\mathrm{nm}$, which allows two-dimensional point-by-point
scanning.

Here, we focus on the noise analysis for the measurement frequency
around 10.4 GHz. The displacement sensitivity is defined as the normalized
minimum detectable vibration amplitude $A_{\textrm{min,total}}^{1/2}=\frac{1}{k}\frac{S_{\textrm{total}}^{1/2}}{\xi R_{\textrm{ref}}}$,
where $R_{\textrm{ref}}$ (around $13\:\textrm{mV}$) is measured
by the LIA and $S_{\textrm{total}}^{1/2}$ is the total noise. $S_{\textrm{total}}^{1/2}$
includes contribution from all available noise sources: electrical
noise (noise without light input) $S_{\textrm{PD}}^{1/2}$ from the
PD, shot noise $S_{\textrm{shot}}^{1/2}$ due to the discrete nature
of photon, and the noise from other electrical parts (including LIA,
amplifier and so on). With the PD connected, and with no incident
light, the PD noise is measured as $S_{\textrm{PD}}^{1/2}=9\:\textrm{nV/\ensuremath{\sqrt{\textrm{Hz}}}}$,
which corresponds to a measurement sensitivity $A_{\textrm{min,PD}}^{\textrm{1/2}}=\frac{1}{k}\frac{S_{\textrm{PD}}^{1/2}}{\xi R_{\textrm{ref}}}=0.34\:\textrm{pm}/\sqrt{\textrm{Hz}}$.
The shot noise is given by $S_{\textrm{shot}}^{1/2}=\frac{1}{2}\times\sqrt{2e\alpha P_{\textrm{in}}B}\times G$,
where $e$ is the elementary charge, $P_{\textrm{in}}$ the incident
laser power on the PD, $B$ the detection bandwidth, and $G$ the
gain of build-in transimpedance amplifier of the PD \citep{ShotNoise}.
In our measurements, $\alpha=0.22\:\textrm{A/W}$ , $P_{\textrm{in}}=200\:\mu\textrm{W}$,
$B=1\:\textrm{Hz}$, and $G=3\:\textrm{V/mA}$. The prefactor $\frac{1}{2}$
is attributed to the voltage division between the $50\:\Omega$ output
impedance of the PD and the $50\:\Omega$ input impedance of the LIA.
Therefore, the resulting shot-noise limited sensitivity becomes $A_{\textrm{min,shot}}^{\textrm{1/2}}=\frac{1}{k}\frac{S_{\textrm{shot}}^{1/2}}{\xi R_{\textrm{ref}}}=0.015\:\textrm{pm}/\sqrt{\textrm{Hz}}$.
The sum of electrical noise from the PD and the shot noise ($\sqrt{A_{\textrm{min,PD}}+A_{\textrm{min,shot}}}$)
gives a theoretical sensitivity limit of $0.34\:\textrm{pm}/\sqrt{\textrm{Hz}}$,
which is close to our measured value $0.36\:\textrm{pm}/\sqrt{\textrm{Hz}}$.
The above analysis shows that the PD noise is the dominant noise source.
For the measurement frequency of 1.67 GHz and 635 MHz, the sensitivity
is around $0.33\:\textrm{pm}/\sqrt{\textrm{Hz}}$, which is consistent
with the above conclusion.

Figure$\,$\ref{fig2}(a) shows the SEM image of the device. A suspended
\emph{c}-axis-oriented AlN microdisk with radius $5.5\:\textrm{\ensuremath{\mu}m}$
and thickness 550 nm is fabricated adjacent to an optical waveguide.
The radius of the supporting pedestal is controlled to be around 100
nm. The photonic structure is patterned using the two-step e-beam
lithography and dry etching technique \citep{Fabdisk,XuNJPpaper,Xu10GHzdisk}.
We actuate the microdisk piezoelectrically using a dipolar pin antenna
and read out the mechanical motion optically through measuring the
optical transmission signal of the optical whispering-gallery mode
(Fig.$\,$\ref{fig2}(a)). This procedure is described in detail in
Ref.$\,$40. Figure$\,$\ref{fig2}(b) shows a broad-band driven
response spectrum. The radial modes can be distinctly identified up
to the 5th-order at 4.5 GHz and the fundamental thickness mode at
10.37 GHz. Zoomed-in spectra of three modes (indicated by the red
stars in Fig.$\,$\ref{fig2}(b)) are shown in Fig.$\,$\ref{fig2}(c).
The quality ($Q$) factors measured are fitted to be around 3200,
3300 and 1800, respectively. 

Figure$\,$\ref{fig2}(d) shows the frequency responses obtained by
our heterodyne interferometer where the probe laser spot is focused
on the center of the microdisk and the signal generator operating
frequency $\omega_{\textrm{mix}}$ is scanned with a fixed electric
excitation of around 25 dBm for the fundamental and 2nd-order radial
modes and 20 dBm for fundamental thickness mode. As shown in Fig.$\,$\ref{fig2}(d),
the vibrometer measurements generally agree with optomechanically-transduced
responses in Fig.$\,$\ref{fig2}(c). Due to reduced out-of-plane
displacement, the third-order and above radial modes are not detected
using our heterodyne interferometer. 

\begin{figure}
\begin{centering}
\includegraphics[clip,width=1\columnwidth]{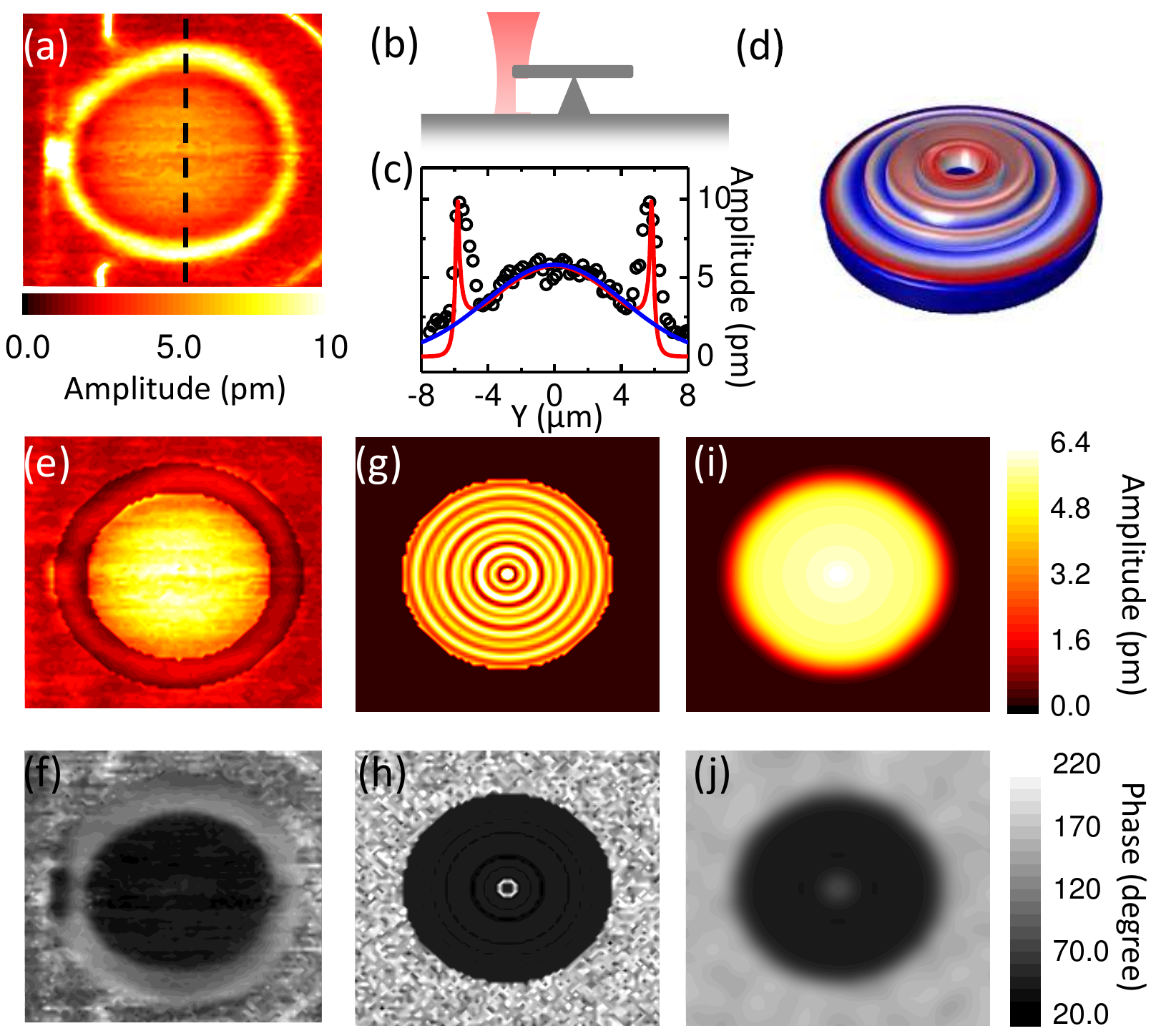}
\par\end{centering}
\caption{(a) Measured amplitude distribution of the thickness mode excited
at a 10.37 GHz drive with a detection bandwidth of 10 Hz. (b) The
laser spot locates at the edge of the microdisk. Due to the light
diffraction, both the AlN mirodisk and silicon substrate contribute
to the reflected light. (c) Black dots indicate the measured amplitude
distribution along the dash black line in (a). The blue line and red
line denote the calculated amplitude distribution before and after
the introduction of the measurement error at the edge of the microdisk.
(d) The simulated pattern of the thickness mode. (e) Error-filtered
amplitude image of (a). (f) Measured phase distribution of the thickness
mode. (g)-(h) The simulated amplitude and phase fields of the thickness
mode, respectively. (i)-(j) The simulated images after convolution
with the laser spot radius of 900 nm. }

\label{fig3}
\end{figure}

The demodulated two-dimensional (2D, $16\times16\:\mu\textrm{m}^{2}$)
vibration amplitude ($Z_{0}$) and phase ($\phi$) data of the fundamental
thickness mode (10.37 GHz) are presented in Figs.$\,$\ref{fig3}(a)
and 3(f). The strong signal at the edge of the microdisk is due to
the multi-path interference of light bouncing back at different sample
heights. As illustrated in Fig.$\,$\ref{fig3}(b), when the laser
spot is fixed at this position, the probe light field became $E_{p}^{\prime}=A_{1}\mathrm{cos}[\omega t+\phi_{1}+2kZ_{0}\mathrm{cos}(\Omega t+\phi)]+A_{3}\mathrm{cos}[\omega t+\phi_{3}]$,
where the $A_{3}$ and $\phi_{3}$ denote the amplitude and phase
of the reflected probe field by the silicon substrate. The demodulated
signal $R_{\textrm{ref}}$ of LIA is determined by the relative phase
$\phi_{3}-\phi_{1}$, as well as $A_{1}$ and $A_{3}$. The phase
depends on the structural parameters (2.2-$\mu\textrm{m}$-thick $\textrm{SiO}_{2}$
buffer layer, 550-nm-thick microdisk)  and $A_{1}$ ($A_{3}$) can
be directly measured in the experiment. Such effect has also been
described in detail by Rembe and Dräbenstedt \citep{VibConfocal}.
As shown by the red line in Fig.$\,$\ref{fig3}(c), the fitting by
considering this edge effect agrees well with the experiment results
(black dots). Figure$\,$\ref{fig3}(e) plots the error-filtered amplitude
image of Fig.$\,$\ref{fig3}(a). Similar analyses are performed for
the other vibration modes discussed later in this paper. Figure$\,$\ref{fig3}(d)
displays the simulated thickness mode shape using the three-dimensional
finite-element method (COMSOL Multiphysics). For the out-of-plane
motion (perpendicular to the surface of microdisk), multi-turn concentric
rings are found in the amplitude distribution field, while the phase
field is uniform except the center where the pillar supports the microdisk,
as shown in Figs.$\,$\ref{fig3}(g) and$\,$\ref{fig3}(h), respectively.
Because of the limited lateral resolution (around 900 nm for our setup),
the spatial images we detect are the convolution transformation of
the vibration fields, as shown in Figs.$\,$\ref{fig3}(i) and$\,$\ref{fig3}(j).
The convolution transformation formula is $Z'(x,y)=0.57\iint dx^{\prime}dy^{\prime}\cdot Z_{sim}(x^{\prime},y^{\prime})\cdot exp(-\frac{(x^{\prime}-x)^{2}+(y^{\prime}-y)^{2}}{r^{2}/2})$,
where $r=900\:\textrm{nm}$ is the radius of the Gaussian-shape spot,
$Z'$ is the expected displacement of measurement, $Z_{sim}$ is the
simulated displacement, and 0.57 is the normalization factor. The
simulated convolution images of amplitude and phase response match
well the experiment results. 

\begin{figure}
\begin{centering}
\includegraphics[clip,width=1\columnwidth]{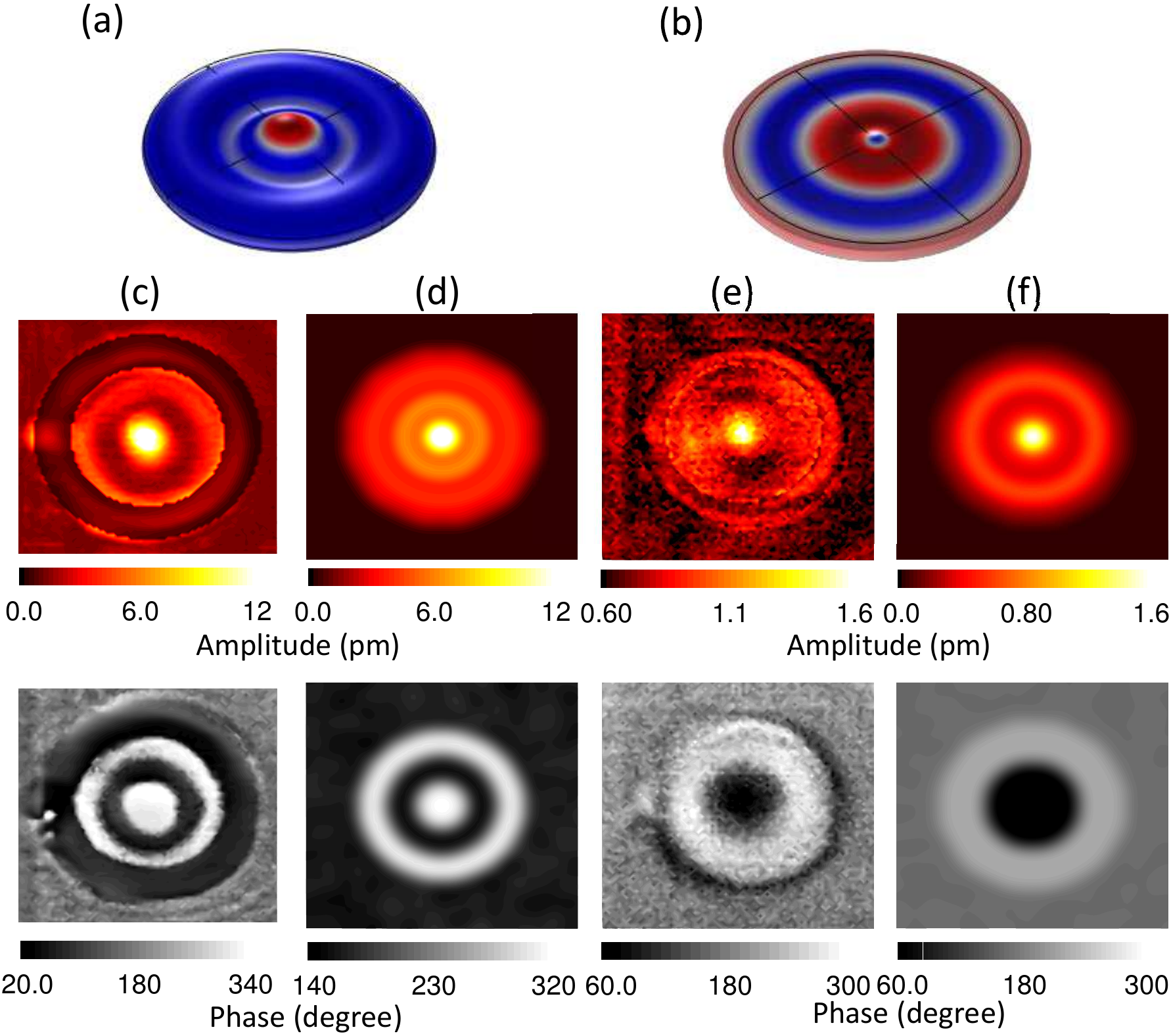}
\par\end{centering}
\caption{(a)-(b) Simulated displacement profiles of the fundamental and the
second-order radial modes, respectively. (c)-(f) The measured amplitude
and phase fields of the fundamental radial mode in (c) and the second-order
radial mode in (e), respectively. The simulated amplitude and phase
fields of the out-of-plane motion for the fundamental mode in (d)
and the second-order radial mode in (f) , respectively.}

\label{fig4}
\end{figure}

Figures$\,$\ref{fig4}(a) and$\,$\ref{fig4}(b) show the simulated
patterns of the fundamental and the second-order radial modes, whose
resonant frequencies are 635 MHz and 1.6 GHz, respectively. The measured
amplitude and phase fields are plotted in Figs.$\,$\ref{fig4}(c)
and$\,$\ref{fig4}(e) with the simulated of out-of-plane motion shown
in Figs.$\,$\ref{fig4}(d) and$\,$\ref{fig4}(f). Benefiting from
the 2D images of acoustic field distribution, especially the phase
information, we can identify the observed modes unambiguously. For
both vibrational modes, the energy of out-of-plane motion concentrates
in the central part of the mcirodisk, we can speculate that the elastic
wave leakage at the pedestal will induce significant mechanical loss
for radial modes. Therefore, reducing the size of the supporting pedestal
can improve the $Q$ factor of the mechanical modes. Simulation results
show that the $Q$ factors are proportional to $\sigma^{-1.1}$ for
the fundamental radial mode and $\sigma^{-2.1}$ for the second-order
radial mode, respectively, where $\sigma$ is the surface area of
the supporting pedestal. 

In conclusion, we have developed a high frequency scanning heterodyne
vibrometer, which can measure the mechanical motion up to 10 GHz.
It not only produces high-resolution amplitude distribution of the
vibration modes but also clearly maps out the phase of the vibration.
This unique capability allows us to fully characterize the thickness
mode and radial modes supported by the AlN microdisk and make accurate
comparisons with numerical simulations. The phase-sensitive heterodyne
vibrometer demonstrated in our paper is a powerful tool for characterizing
mechanical modes in micromechanical components, and enables better
design and performance of high frequency optomechanical and electromechanical
devices.

\textbf{Acknowledgment} This work is supported by DARPA/MTO's PRIGM:
AIMS program through a grant from SPAWAR (N66001-16-1-4026), the Laboratory
of Physical Sciences through a grant from Army Research Office (W911NF-14-1-0563)),
an Air Force Office of Scientific Research (AFOSR) MURI grant (FA9550-15-1-0029)
and a NSF MRSEC grant (1119826). H.X.T. acknowledges support from
a Packard Fellowship in Sceince and Engineering. The authors thank
Michael Power and Dr. Michael Rooks for assistance in device fabrication.


\begin{thebibliography}{99}
\bibitem{Appli1}M. Aspelmeyer, T. J. Kippenberg, and F. Marquardt,
Rev. Mod. Phys. \textbf{86}, 1391 (2014).

\bibitem{Appli2}C. A. Regal and K. W. Lehnert, J. Phys. Conf. Ser.
\textbf{264}, 012025 (2011).

\bibitem{Appli3}R. W. Andrews, R. W. Peterson, T. P. Purdy, K. Cicak,
R. W. Simmonds, C. A. Regal, and K. W. Lehnert, Nat. Phys. \textbf{10},
321 (2014).

\bibitem{Cool1}J. D. Teufel, T. Donner, D. Li, J. W. Harlow, M. S.
Allman, K. Cicak, A. J. Sirois, J. D. Whittaker, K. W. Lehnert, and
R. W. Simmonds, Nature (London) \textbf{475}, 359 (2011).

\bibitem{Cool2}J. Chan, T. P. M. Alegre, A. H. Safavi-Naeini, J.
T. Hill, A. Krause, S. Gröblacher, M. Aspelmeyer, and O. Painter,
Nature (London) \textbf{478}, 89 (2011).

\bibitem{Cool3}R. W. Peterson, T. P. Purdy, N. S. Kampel, R. W. Andrews,
P.-L. Yu, K. W. Lehnert, and C. A. Regal, Phys. Rev. Lett. \textbf{116},
063601 (2016).

\bibitem{strong1}J. D. Teufel, D. Li, M. S. Allman, K. Cicak, a.
J. Sirois, J. D. Whittaker, and R. W. Simmonds, Nature (London) \textbf{471},
204 (2011).

\bibitem{strong2}E. Verhagen, S. Deléglise, S. Weis, A. Schliesser,
and T. J. Kippenberg, Nature (London) \textbf{482}, 63 (2012).

\bibitem{strong3}X. Han, C. L. Zou, and H. X. Tang, Phys. Rev. Lett.
\textbf{117}, 123603 (2016).

\bibitem{Conversion1}C.-H. Dong, V. Fiore, M. C. Kuzyk, and H.-L.
Wang, Science \textbf{338}, 1609 (2012).

\bibitem{Conversion2}T. Bagci, A. Simonsen, S. Schmid, L. G. Villanueva,
E. Zeuthen, J. Appel, J. M. Taylor, A. Sørensen, K. Usami, A. Schliesser,
and E. S. Polzik, Nature (London) \textbf{507}, 81 (2014).

\bibitem{Nonreciprocity1}C. H. Dong, Z. Shen, C. L. Zou, Y.L. Zhang,
W. Fu, and G. C. Guo, Nat. Commun. \textbf{6}, 6193 (2015).

\bibitem{Nonreciprocity2}W. Fu, F. J. Shu, Y. L. Zhang, C. H. Dong,
C. L. Zou, and G. C. Guo, Optics Express \textbf{23}, 25118-225127
(2015).

\bibitem{Nonreciprocity3}Z. Shen, Y. L. Zhang, Y. Chen, C. L. Zou,
Y. F. Xiao, X. B. Zou, F. W. Sun, G. C. Guo, and C. H. Dong, Nat.
Photon. \textbf{10}, 657 (2016). 

\bibitem{10GHzImprotant}A. D. O\textquoteright Connell, M. Hofheinz,
M. Ansmann, R. C. Bialczak, M. Lenander, E. Lucero, M. Neeley, D.
Sank, H. Wang, M. Weides, J. Wenner, J. M. Martinis, and A. N. Cleland,
Nature \textbf{464}, 697 (2010).

\bibitem{book1}C. Campbell, \emph{Surface Acoustic Wave Devices and
their Signal Processing Applications} (Academic, 1989).

\bibitem{communication1}Y. Satoh, T. Nishihara, T. Yokoyama, M. Ueda,
and T. Miyashita, Jpn. J. Appl. Phys. \textbf{44}, 2883 (2005). 

\bibitem{MoLi1}S. A. Tadesse, and M. Li, Nat. Commun. \textbf{5},
5402 (2014).

\bibitem{MoLi2}H. Li, S. A. Tadesse, Q. Liu, and M. Li, Optica \textbf{2},
826-831 (2015).

\bibitem{Vib1SAW}J. V. Knuuttila, P. T. Tikka, and M. M. Salomaa,
Opt. Lett. \textbf{25}, 613\textendash 615 (2000).

\bibitem{Vib2thin}J. E. Graebner, B. P. Barber, P. L. Gammel, D.
S. Greywall, and S. Gopani, Appl. Phys. Lett. \textbf{78}, 159\textendash 161
(2001).

\bibitem{Vib3thin}G. G. Fattinger and P. T. Tikka, Appl. Phys. Lett.
\textbf{79}, 290\textendash 292 (2001).

\bibitem{Timeresolved1}T. Tachizaki, T. Muroya, O. Matsuda, Y. Sugawara,
D. H. Hurley, and O. B. Wright, Rev. Sci. Instrum. \textbf{77}, 043713
(2006).

\bibitem{VibConfocal}C. Rembe, and A. Dräbenstedt, Rev. Sci. Instrum.
\textbf{77}, 083702 (2006).

\bibitem{Vib4Canti}P. Sanz, J. Hernando, J. Vazquez, and J. L. Sanchez-Rojas,
J. Micromech. Microeng. \textbf{17}, 931-937 (2007).

\bibitem{Vib5drump}H. Martinussen, A. Aksnes, and H. E. Engan, Opt.
Express \textbf{15,} 11370-11384 (2007).

\bibitem{Vib6thin}K. Kokkonen, and M. Kaivola, Appl. Phys. Lett.
\textbf{92,} 063502 (2008).

\bibitem{Timeresolved2}T. Fujikura, O. Matsuda, D. M. Profunser,
O. B. Wright, J. Masson, and S. Ballandras, Appl. Phys. Lett. \textbf{93,}
261101 (2008).

\bibitem{Vib7Canti}L. C. Chen, Y. T. Huang, X. L. Nguyen, J. L. Chen,
and C. C. Chang, Opt. Lasers Eng. \textbf{47} 237-251 (2009).

\bibitem{Vib8thin}K. Y. Hashimoto, K. Kashiwa, N. Wu, T. Omori, M.
Yamaguchi, O. Takano, S. Meguro, and K. Akahane, IEEE Trans. Ultrason.
Ferroelect. Freq. Control \textbf{58}, 187-194 (2011).

\bibitem{Vib9drump}E. Leirset, H. E. Engan, and A. Aksnes, Opt. Express\textbf{
21,} 19900-19921 (2013).

\bibitem{VibSiC}Z. Wang, J. Lee, and P. X.-L. Feng, Nat. Commun.
\textbf{5}, 5158 (2014) 

\bibitem{Vibwaveguide}Z. Shen, W. Fu, R.-S. Cheng, H. Townley, C.
L. Zou, and H. X. Tang, `` High frequency scanning vibrometer for
phase sensitive visualization of guided surface acoustic wave modes,''
in preparation.

\bibitem{Vibring}W. Fu, Z. Shen, C. L. Zou, and H. X. Tong, ``High-Q
Surface acoustic wave ring resonator on a chip,'' in preparation. 

\bibitem{Timeresolved3}T. A. Kelf, Y. Tanaka, O. Matsuda, E. M. Larsson,
D. S. Sutherland and O. B. Wright, Nano Lett. \textbf{11}, 3893-3898
(2011).

\bibitem{Timeresolved4}S. Mezil, P. H. Otsuka, S. Kaneko, O. B. Wright,
M. Tomoda, and O. Matsuda, Opt. Lett. \textbf{40}, 2157-2160 (2015).

\bibitem{ShotNoise}R. L. Whitman, and A. Korpel, Appl. Opt. \textbf{8},
1567-1576 (1969). 

\bibitem{Fabdisk}C. Xiong, X. Sun, K. Y. Fong, and H. X. Tang, Appl.
Phys. Lett. \textbf{100}, 171111 (2012).

\bibitem{XuNJPpaper}X. Han, C. Xiong, K. Y. Fong, X. F. Zhang, and
H. X. Tang, New J. Phys. \textbf{16, }063060 (2014)\textbf{.}

\bibitem{Xu10GHzdisk}X. Han, K. Y. Fong, and H. T. Tang, Appl. Phys.
Lett. \textbf{106}, 161108 (2015).
\end{thebibliography}
\end{document}